# Origin of the increased velocities of domain wall motions in soft magnetic thin-film nanostripes beyond the velocity-breakdown regime

Sang-Koog Kim,[*] Jun-Young Lee, Youn-Seok Choi, Konstantin Yu. Guslienko, and Ki-Suk Lee

*Research Center for Spin Dynamics & Spin-Wave Devices, Seoul National University*

*Nanospinics Laboratory, Department of Materials Science and Engineering, College of Engineering, Seoul National University, Seoul 151-744, Republic of Korea*

It is known that oscillatory domain-wall (DW) motions in soft magnetic thin-film nanostripes above the Walker critical field lead to a remarkable reduction in the average DW velocities. In a much-higher-field region *beyond the velocity-breakdown regime*, however, the DW velocities have been found to increase in response to a further increase of the applied field. We report on the physical origin and detailed mechanism of this unexpected behavior. We associate the mechanism with the serial dynamic processes of the nucleation of vortex-antivortex (V-AV) pairs inside the stripe or at its edges, the non-linear gyrotropic motions of Vs and AVs, and their annihilation process. The present results imply that a two-dimensional soliton model is required for adequate interpretation of DW motions in the linear- and oscillatory-DW-motion regimes as well as in the beyond-velocity-breakdown regime.

Magnetic field-driven magnetization (**M**) reversals in bulk or thin-film magnetic materials via domain-wall (DW) movements as well as coherent or incoherent **M** rotation have been historically longstanding issues in the research field of magnetism.[1] The motion of a head-to-head (or tail-to-tail) DW in magnetic thin-film nanostripes under applied magnetic fields and/or spin-polarized current, the subject of cutting edge research,[2-10] has very promising potential applications to new classes of race-track memory[11] and logic[12] devices. When a static field, $H$, is applied to a DW, its velocity increases with increasing $H$ in a low-$H$ region, but remarkably drops above a certain critical field known as the Walker field, $H_w$.[2-4,13-15] From earlier studies[16, 17] it is known that in the low-$H$ region a single transverse wall (TW) moves steadily along the nanostripe parallel to the field direction. The velocity breakdown is known to be associated with the oscillations of the internal structure of a single moving DW between the TW and the vortex or antivortex wall (VW or AVW)[16] with the time period of $T = 2\pi/\gamma H$,[17] where $\gamma$ is the gyromagnetic ratio. In the velocity-breakdown region, V or AV within the VW or AVW exhibits a non-linear gyrotropic motion in the nanostripe, resulting in the reduced average DW velocity due to the V or AV motion against the **M** reversal direction (backward or toward the $H$ direction). In a much-higher-field region beyond the velocity-breakdown regime, on the contrary the average DW velocity has been typically observed to again increase with a further increase of the applied field.[4,16] The physical origin and detailed understanding of this unexpected behavior remains elusive, whereas the linear and oscillatory DW dynamics have been described by one-dimensional (1D) models[13,18] and a 2D dynamic soliton model.[16,17] The dynamic origin of the increased DW velocities in the higher-$H$ region beyond the DW oscillatory regime is as yet unknown.

In this article, we report on the underlying mechanism of **M** reversal dynamics in soft magnetic thin-film nanostripes beyond the velocity-breakdown regime, which occurs via the serial processes of the nucleation, gyrotropic motion, and annihilation of V-AV pairs, while maintaining the conservation of their total topological charge in the nanostripes. The present results offer a new and deeper physical insight in V-AV-mediated **M** reversal dynamics in nanostripes *beyond the velocity-breakdown regime*. We confirm that a 2D dynamic soliton model proposed in our earlier work[16,17] is necessary for an adequate physical description of **M** reversal dynamics, covering all the different field regimes of DW motions in nanostripes.

In the present micromagnetic simulations, we used a rectangular-shaped soft magnetic nanostripe (made of Permalloy, Py) of 10 nm thickness ($L$), 6 $\mu$m length, and different width, $w =$ 140, 240 and, 400 nm. The simulation procedure and details were reported in Ref. 16. The initial DW type used in this study was a head-to-head TW having non-zero average transverse **M** as the ground state at zero field in the given nanostripe [Fig. 1(a)]. This DW was driven by a static magnetic field of varying strength applied along the long axis of the nanostripes (in the $+x$ direction).

The **M** reversal dynamics for $w = 140$ nm and with three different fields, $H = 5, 25,$ and 150 Oe, are represented by the displacements of DW versus time ($D$-vs-$t$) curves shown in Fig. 1(b). Each of the three different field strengths was selected from each of the represented three distinct $H$ regions, noted as $H < H_W$, $H_W < H < H_{mult}$, and $H > H_{mult}$, hereafter simply noted as Regions I, II, and III, as reported in Ref. 16. $H_{mult}$ is the field by which the oscillatory DW dynamics above $H_W$ are distinguished from the region where a multi-V-AV state appears during the **M** reversal. In this geometry, $H_W$ and $H_{mult}$ were determined to be approximately 10 and 90 Oe, respectively. Those characteristic *D*-vs-*t* curves indicate that the speeds of the **M** reversals are quite different. In Region III ($H$ = 150 Oe), that is, well above the velocity-breakdown regime, the **M** reversal again becomes faster (the average DW velocity $\bar{\upsilon}$ =162 m/s) than that in Region II ($\bar{\upsilon}$ = 66 m/s) as well as more or less comparable to that in Region I ($\bar{\upsilon}$ = 231 m/s). As another example, the *D*-vs-*t* for $w$ =240 nm and $H$ = 100 Oe is also shown, indicating that the reduced velocity of DW motions in Region II is markedly increased in Region III. This reflects the fact that the underlying reversal mechanisms in the individual field regions are quite distinct. For the details of other regions I and II, see Refs. 16, 17.

To understand the exact **M** reversal mechanism in Region III, the serial snapshot images shown in Fig 2(a) were taken at the indicated times to represent the temporal evolution of the head-to-head DW for $w$ = 140 nm and with $H$ = 150 Oe. The similar images shown in Fig 2(b) for $w$ = 240 nm and with $H$ = 100 Oe illustrate a different reversal mechanism, as will be discussed below. Unlike the motions in Regions I and II,[16,17] the DW motion in Region III is neither a steady motion of a TW nor a periodic oscillation of the internal structure of a *single* moving DW between its different types (TW, VW, or AVW). Rather, the internal structure of the moving DW contains several magnetic topological solitons[16,17] such as Vs and AVs whose cores are seen as black or white spots according to the **M** orientation of their cores (polarization, $p$ ; $p$ = +1(-1) for the up-(down-)core orientation. Those serial snapshot images display the individual steps associated with the V-AV-mediated dynamic processes. To elucidate the characteristic dynamic features of the nucleation, gyrotropic motion, and annihilation of the V-AV pairs, in Fig 3 we presented a schematic illustration and the trajectories of the cores of Vs and AVs during their gyrotropic motions in the plane of a given nanostripe.

Case I, shown in Fig. 3(a), illustrates that the V and AV with the same up-core orientation nucleate (①) at the opposite edges of the nanostripe at nearly the same time. Then, both the V and AV move inward the nanostripe, become closed (②), according to their own non-linear gyrotropic motions, and then finally collapse (③) inside the nanostripe and annihilate [see their core trajectories (right panel)]. Immediately after the V-AV annihilation process, strong spin waves are emitted (③) according to the mechanism reported earlier.[19] For convenience, hereafter, we denote Vs and AVs by the symbols $V$ and $\overline{V}$, respectively, and their core polarizations by the arrows ↑ and ↓ in the symbols' subscripts, respectively. On the other hand, case II, for $w = 240$ nm and $H = 100$ Oe,[20] is shown in Fig. 3(b), where $V_\uparrow$ and $\overline{V}_\uparrow$ simultaneously nucleate (①) at the opposite edges as in case I. However, unlike case I, the $V_\uparrow$ moves (②) inward faster than $\overline{V}_\uparrow$, while the $\overline{V}_\uparrow$ is maintained near the same stripe edge. As the next step, the $V_\uparrow$ moves further inward, and around it an additional $V'_\downarrow$ - $\overline{V}'_\downarrow$ pair with $p = -1$ is nucleated (③). The prime in the superscripts is used to distinguish the newly nucleated V-AV pair from the V-AV pair ($V_\uparrow$ - $\overline{V}_\uparrow$) nucleated earlier at both nanostripe edges. Then, the $V_\uparrow - \overline{V}'_\downarrow$ pair annihilation process is followed (④) according to the mechanism reported in Refs.16, 21, and is accompanied by spin wave emission by the core reversal.[19] The remaining new vortex ($V'_\downarrow$) with the reversed core polarization moves towards the up edge of the nanostripe according to its own gyrotropic motion (⑤), as illustrated in the schematic drawing (middle panel) and serial snapshot images, as well as in the core trajectories (right panel) in Fig. 3. Although cases I and II differ in the overall **M** reversal mechanism, the same serial processes of the individual steps of the nucleation, non-linear gyrotropic motions, and annihilation of the Vs and AVs commonly apply.

In order to gain better physical insight into those common individual processes, it is necessary to adopt the V-AV-mediated DW dynamics with regard to the emission, motion, and absorption of topological solitons in the framework of collective dynamic variables – the soliton core positions. The in-plane **M** configuration of a single symmetric TW resembles an isosceles triangle with three apexes [Fig. 1(a), right]. An AV can nucleate only at a single apex, in this case, placed at the down stripe edge, and a V can nucleate at one of the double apexes at the upper edge, so that the three apexes can act as the nucleation sites for the magnetic solitons. This is so, because the **M** configuration of the moving TW can be represented by magnetic solitons of half-integer topological charges $q$ located on the stripe edges, as shown in the right of Fig. 1(a). A similar approach was developed to explain the oscillatory transformation of the internal structure of the different-type DWs in the nanostripes within the velocity-breakdown regime.[16] In Region II, oscillatory DW motions are associated with the transformation of the internal structure of a single moving DW between a TW and a V or an AV.

In the higher-field Region III, however, both V and AV simultaneously nucleated at the opposite edges, as seen in Fig. 3. Such nucleations, compared with that of a single V or AV in Region II, lead to the increase of the exchange energy due to the presence of additional soliton cores. These events occur owing to a remarkable reduction of the Zeeman energy by a further increase of $H$, resulting in the excess exchange energy, due to the increased number of the soliton cores inside the nanostripe, being compensated. For the cases of further increases of $w$ and/or $H$, the number of the nucleated V-AV pairs was observed to be increased. This **M** reversal is much faster than those via a single V or AV motion or a single V-AV pair mediated DW motion. Consequently, the reason for the increased DW velocities in the beyond-velocity-breakdown regime is because V-AV-pair nucleation is the dominant reversal process in such high-$H$ region. In such V-AV-mediated **M** dynamics, the total topological charge of all magnetic solitons (indexed by $j$) inside the nanostripe should be maintained constant, i.e., $\sum_j q_j = const$. For the case of this nanostripe geometry, the constant value is zero because the TW initially has half integer charges, $q = +1/2$ and $-1/2$ at the opposite edges. The V and AV also bear an integer topological charge, $q = 1$ and $q = -1$, respectively.[17]

Next, to understand the observed complex trajectories of the coupled V and AV motions shown on the right side of Fig. 3, it is necessary to introduce the gyrovector, $\mathbf{G} = -Gpq\hat{\mathbf{z}}$,[22] where $\hat{\mathbf{z}}$ is the unit vector perpendicular to the stripe plane. Since $G = 2\pi M_s L/\gamma$ with $M_s = |\mathbf{M}|$ is a positive constant for the given **M** distribution,[23] the non-zero gyrovector of a soliton results in the gyrotropic motion of the soliton center position, **X**, in a given potential profile $W$. Additionally, the magnetostatic energy plays an important role in forming the potential well (hill) for a single V(AV) with respect to the middle of the nanostripe.[16] For the present nanostripe geometry, the sense of the gyromotion of a single V or AV is thus determined by the sign of the stiffness coefficient $\kappa = |\kappa|\text{sign}(q)$ of the corresponding $W$ in the transverse direction as well as by the product of $pq$. Therefore, the direction of a single AV or V motion in Region II depends only on the soliton polarization $p$.[16,17] However, having considered interactions between V and AV or the same types, such as interacting Vs and AVs, the potential profile is not simply determined by the sign of $q$ as in $\kappa = |\kappa|\text{sign}(q)$. For instance, the core trajectories of $V_\uparrow$ and $\bar{V}_\uparrow$ in Fig. 3(a), are CCW rotation for a certain period of time after their nucleation at stripe edges, but $\bar{V}_\uparrow$ changes its rotation sense to CW owing to the attractive interaction between $V_\uparrow$ and $\bar{V}_\uparrow$ when they become closed near the stripe center. The exchange interaction energy of the two solitons with the charges $q_1$, $q_2$ and the core positions $\mathbf{X}_1, \mathbf{X}_2$ can be written as $W_{int} = -4\pi A L q_1 q_2 \ln(|\mathbf{X}_1 - \mathbf{X}_2|/L_e)$ with $L_e = \sqrt{2A}/M_s$ and exchange stiffness $A$, which is relatively small in comparison with the magnetostatic interaction.[24] V and AV attract each other, and the solitons with the same sign of $q$ repel each other, independently of the $p$ and chirality of V and AV (the sense of the in-plane **M** orientation around its core). A more complex example is shown in Fig. 3(b), where $V_\uparrow$ and $V'_\downarrow$ seem to follow the rotation senses determined by their $p$'s, but their motion directions with respect to the field direction are different from the motions in the oscillatory transformation of a single moving DW, as reported in Ref. 16. Note that the DW velocity in Region III calculated in Ref. 25 is too small to explain the experimental results.[4,8,9] Further studies on the effect of the exchange and magnetostatic interactions between V and AV on their gyrotropic motions are necessary.

According to the results shown above, we conclude the following. In Region I, the **M**-reversal velocity is determined by the velocity of a single TW motion, which strongly depends on the damping parameter and the applied field strength, and in Region II, the reversal velocity is determined by the repeated forward and backward motion of a single V or AV that was nucleated at the stripe edges and transformed from a TW, resulting in the reduced average velocity. Meanwhile, in Region III beyond the velocity-breakdown regime, the **M**-reversal dynamics are a consequence of the serial dynamic processes of the nucleation, gyrotropic motion, and annihilation of coupled Vs and AVs inside or at the edges of nanostripes, which processes repeatedly take place to complete the **M** reversal. Thereby, the **M-**reversal velocity in Region III is determined by the nucleation process of V-AV pairs, which is the dominant reversal process, followed by the gyrotropic motions of the individual V and AV and their annihilation. Such nucleation of V-AV pairs can occur at several local regions for further increase of *H*, in order to increase the number of V-AV pairs, making the **M** reversal much faster than that in Region II. Nucleation of the coupled V-AV pairs, their motions, and annihilation play a crucial role in **M**-reversal dynamics in restricted geometry.

The present results reflect the fact that even though 1D models can, albeit incompletely, explain linear and oscillatory DW motions in low- and intermediate-field regions, they cannot adequately describe the nucleation, gyrotropic motion, and annihilation of V-AV pairs inside nanostripes at high fields. For adequate physical interpretations of DW motions and transformations of the internal DW structure of different-type DWs (TWs, VWs and AVWs) in Regions I and II, as well as of the individual processes of the nucleation, non-linear gyrotropic motion, and annihilation of topological magnetic solitons in soft magnetic thin-film nanostripes, a 2D soliton model should be applied.

This work was supported by Creative Research Initiatives (ReC-SDSW) of MOST/KOSEF.

# References

Corresponding author; sangkoog@snu.ac.kr

[1] A. Hubert, and R. Schafer, *Magnetic Domains* (Springer, Berlin, 2000).

[2] Y. Nakatani, A. Thiaville, and J. Miltat, Nat. Mater. **2**, 521 (2003).

[3] D. Atkinson, D. A. Allwood, G. Xiong, M. D. Cooke, C. C. Faulkner, and R. P. Cowburn, Nat. Mater. **2**, 85 (2003).

[4] G. S. D. Beach, C. Nistor, C. Knutson, M. Tsoi, and J. L. Erskine, Nat. Mater. **4**, 741 (2005); Phys. Rev. Lett. **97**, 057203 (2006).

[5] L. Thomas, M. Hayashi, X. Jiang, R. Moriya, C. Rettner, and S. S. P. Parkin, Nature **443**, 197 (2006); L. Thomas, C. Rettner, M. Hayashi, M. G. Samant, S. S. P. Parkin, A. Doran, and A. Scholl, Appl. Phys. Lett. **87**, 262501 (2005).

[6] .Meier G, Bolte M, and René Eiselt, Phys. Rev. Lett. **98**, 187202 (2007)

[7] J.-Y. Lee, K.-S. Lee, and S.-K. Kim, Appl. Phys. Lett. **91**, 122513 (2007).

[8] G. S. D. Beach, M. Tsoi and J.L. Erskine, J. Magn. Magn. Mater. **320**, 1272 (2008).

[9] J. Yang, C. Nistor, G. S. D. Beach, and J. L. Erskine, Phys. Rev. B. **77,** 014413 (2008).

[10] M. Hayashi, L. Thomas, R. Moriya, C. Rettner, and S. S. P. Parkin, Science **11,** 209-211 (2008); M. Hayashi, L. Tomas, C. Rettner, R. Moriya, and S. S. P. Parkin, Nat. Phys. **3**, 21 (2007); Appl. Phys. Lett. **92**, 112510 (2008).

[11] S. S. P. Parkin, U.S. Patents 6,834,005, 6,898,132, 6,920,062, 7,031,178, and 7,236,386 (2004 to 2007); S. S. P. Parkin, M. Hayashi, and L. Thomas, Science **320**, 190 (2008)

[12] D. A. Allwood, G. Xiong, C. C. Faulkner, D. Atkinson, D. Petit, and R. P. Cowburn, Science **309**, 1688 (2005).

[13] N. L. Schryer and L. R. Walker, J. Appl. Phys. **45**, 5406 (1974).

[14] Y. Nakatani, A. Thiaville, and J. Miltat, J. Magn. Magn. Mater. **290-291**, 750 (2005);

[15] S. W. Yuan and H. N. Bertram, Phys. Rev. B **44**, 12395 (1991).

[16] J.-Y. Lee, K.-S. Lee, S. Choi, K. Y. Guslienko, and S.-K. Kim, Phys. Rev. B **76**, 184408 (2007).

[17] K. Y. Guslienko, J.-Y. Lee, and S.-K. Kim, e-print arXiv:cond-mat/0711.3680.

[18] A. Thiaville, Y. Nakatani, J. Miltat, and N. Vernier, J. Appl. Phys. **95**, 7049 (2004).

[19] S. Choi, K.-S. Lee, K. Y. Guslienko, and S.-K. Kim, Phys. Rev. Lett. **98**, 087205 (2007).

[20] In this case we used the cell size of $3\times3\times10$ nm$^3$ in order to clearly investigate the nucleation of a V-AV pair near the V that was moved from the edge after its nucleation.

[21] B. Van Waeyenberge et al., Nature **444**, 461 (2006).

[22] A. A. Thiele, J. Appl. Phys. **45**, 377 (1974); D. L. Huber, Phys. Rev. B **26**, 3758 (1982).

[23] K. Y. Guslienko, B. A. Ivanov, V. Novosad, Y. Otani, H. Shima, and K. Fukamichi, J. Appl. Phys. **91**, 8037 (2002).

[24] K.S. Buchanan, P. Roy, M. Grimsditch, F. Fradin, K.Y. Guslienko, S.D. Bader, and V. Novosad V., Nat. Phys. **1**, 172 (2005).

[25] O.A. Tretiakov, D. Clarke, G.-W. Chern, Ya. B. Bazaliy, and O. Tchernyshyov, Phys. Rev. Lett. **100**, 127204 (2008).

**Figure captions**

Fig. 1. (Color online) (a) Left: In-plane **M** configuration of a TW with the nonzero transverse (+ y direction) **M** component in nanostripe. The colors denote the local in-plane **M**, as indicated by the color wheel. Right: Representation of half-integer topological charges located at both stripe edges, along with the nucleation sites for V (dark yellow dots) and AV (green dot). The streamlines indicate the in-plane orientation of the local **Ms.** (b) DW displacement versus time curves, for DW motions driven by magnetic fields $H$ along long axis of nanostripe of 10 nm thickness, 140 nm width, and 6 μm length, as reported in Ref. 16.

Fig. 2. (Color online) Serial snapshot images of temporal dynamic evolution of DW motions in a given nanostripe of $w = 140$ nm and with $H = 150$ Oe in (a) and $w = 240$ nm and $H = 100$ Oe. in (b). The gray scale and streamlines indicate the normalized out-of-plane **M** components, as noted by the gray bar, and the in-plane **M** orientation, respectively. The white and black spots indicate the core polarization $p = +1$ and $p = -1$, respectively, of the Vs and AVs. The positions of the moving DWs at indicated times, are noted by numbered images.

Fig. 3. (Color online) Snapshot images (left) and schematic illustrations (middle), representing nucleation, gyrotoropic motion, and annihilation of Vs and AVs, along with spin-wave radiation, (a) and (b) correspond to the dynamic evolutions marked by the red boxes in Fig. 2(a) and 2(b), respectively. Right columns display trajectories of motions of V and AV cores. The different colors denote the different types of vortices, with the indicated core orientations. The arrows nearby the trajectories indicate the directions of the movement of each soliton. The open circles with the numbers on the trajectories describe the core positions at the indicated times noted by the same numbers as those in the plane-view images.

# Figures

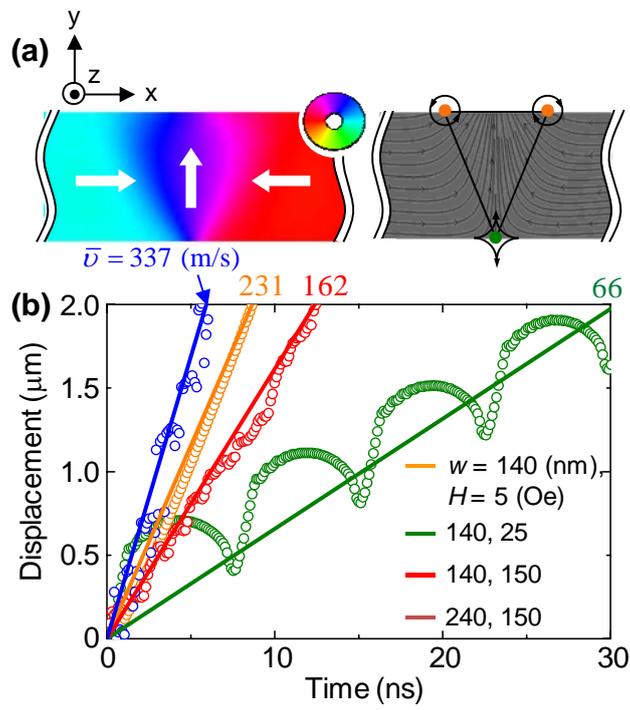

**Fig. 1**

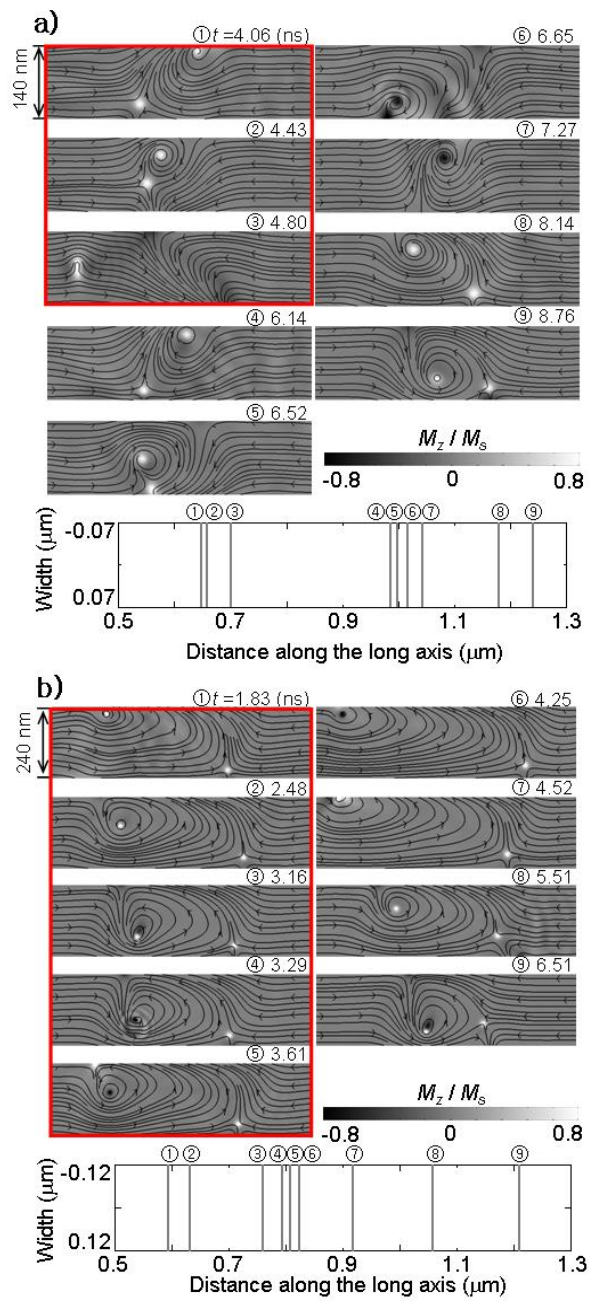

**Fig. 2**

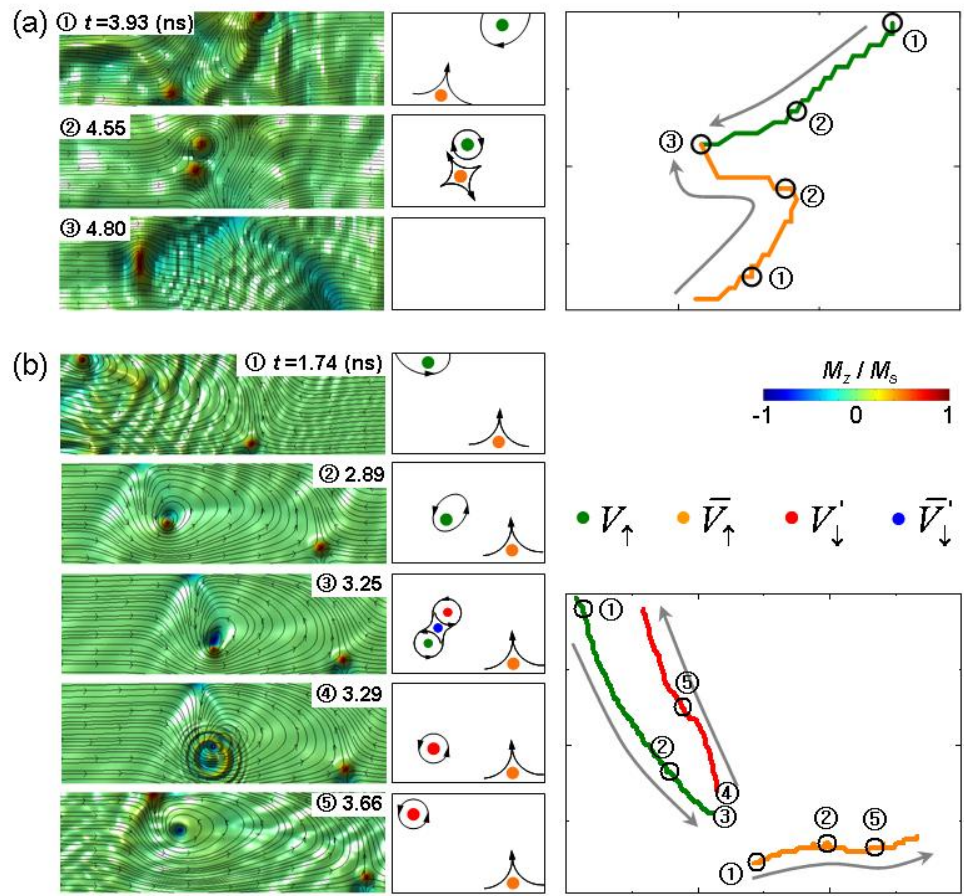

**Fig. 3**